\documentclass[12pt]{iopart}
\usepackage{iopams}
\usepackage{graphicx}  
\begin{document}

\title[LISA Selection Effects]{Selection effects in resolving Galactic binaries with LISA}

\author{M J Benacquista$^1$, S L Larson$^2$, B E Taylor$^3$}

\address{$^1$ Center for Gravitational Wave Astronomy, University of Texas at Brownsville, Brownsville, TX 78520, US}
\address{$^2$ Dept.\ of Phys., Weber State University, 2508 University of Circle, Ogden, UT 84408, US}
\address{$^3$ Dept.\ of Chem. \& Phys., Radford University, Radford, VA 24142, US}\ead{benacquista@phys.utb.edu}
\begin{abstract}
Using several realisations of the Galactic population of close white dwarf binaries, we have explored the selection bias for resolved binaries in the LISA data stream. We have assumed a data analysis routine that is capable of identifying binaries that have a signal to noise ratio of at least 5 above a confusion foreground of unresolved binaries. The resolved population of binaries is separated into a subpopulation over 1000 binaries that have a measureable chirp and another subpopulation over 20,000 binaries that do not. As expected, the population of chirping binaries is heavily skewed toward high frequency, high chirp mass systems, with little or no preference for nearby systems. The population of non-chirping binaries is still biased toward frequencies above about 1 mHz. There is an overabundance of higher mass systems than is present in the complete Galactic population.
\end{abstract}


\section{Introduction}
One of the important science objectives of LISA is to survey compact stellar-mass binaries and study the structure of the Galaxy. One way that this can be accomplished is by measuring the properties of the confusion-limited foreground of signals below about 3 mHz and cataloguing individually resolved binaries. In this paper, we explore the properties of the binaries that are expected to be individually resolved and compare them with the underlying Galactic population. We approach this problem by first synthesising a number of realisations of the Galactic white dwarf binary population using different Galactic structure parameters. Next, we apply a set of detection criteria to this population in order to simulate the success of future data analysis to resolve individual binaries. Finally, we explore the properties of the resolved binaries. We find that the resolved binaries tend to be from the high-mass, high-frequency end of the population. Since these types of binaries are rare, we find that the spatial distribution of the resolved systems is relatively complete throughout the Galaxy. This is particularly true with the resolved systems whose chirp ($\dot{f}$) is due solely to gravitational wave emission and is measurable. For systems that are effectively monochromatic throughout the observation time, we find a slight excess of systems that are nearby, but again the bulk of the distribution of binaries samples the entire Galaxy.

\section{The Galaxy Simulation}
We use the population synthesis of Benacquista \etal~\cite{benacquista04}, which is based on the detailed binary evolution work of Nelemans \etal~\cite{nelemans01}. In our synthesis, the stellar birthrate is assumed to be constant and if a binary evolves to the point that Roche lobe overflow begins, it is removed from the population. Therefore, our model only includes detached systems. We have also assumed that there is no tidal coupling prior to the onset of Roche lobe overflow. Consequently, any frequency evolution of the binaries is solely due to the emission of gravitational radiation. The spatial distribution of the binaries is assumed to be governed by the distribution model:
\begin{equation}
\rho({\bi r}) = \frac{N}{4\pi R R_0 z_0}\rme^{-R/R_0}{\rm sech}^2\left(z/z_0\right),
\end{equation}
where $N$ is the total number of binaries, $R$ and $z$ are the Galactocentric cylindrical co-ordinates and $R_0$ and $z_0$ are the radial and vertical scales of the Galaxy. We note that this density distribution has a factor of $R$ in the denominator, which provides a reasonable approximation of a bulge population~\cite{nelemans03}.

We have varied the radial scale and the vertical scale in order to explore the effect of Galactic structure on the population of resolved binaries. (The reverse problem of determining Galactic structure from the population of resolved binaries will be addressed in another paper~\cite{larson07}.) We have chosen $R_0 = 2.0,~2.5,~3.0$ kpc and $z_0 = 100,~200,~300,~500$ pc for our models. Our baseline model (model C) has $R_0= 2.5$ kpc and $z_o = 200$ pc. We have used the calibration of Hils \etal~\cite{hils90}, which is based on the surface density star formation rate to obtain $N = 3 \times 10^7$. We have used the same calibration on all our models except for model G, where we have assumed the same local space density as found in our baseline model. This results in a value of $N = 7.5 \times 10^7$ for model G. For each model, we have generated 3 realisations of the white dwarf binary population. The seven Galactic structure models we have used are given in Table~\ref{galaxymodels}.
\begin{table}
\caption{\label{galaxymodels} Parameter choices for different structure models of the Galaxy. The baseline model has $R_0 = 2.5$ kpc, $z_0 = 200$ pc, and $N = 3 \times 10^7$.}
\begin{indented}
\item[]\begin{tabular}{clll}
\br
Model & $R_0$ (kpc) & $z_0$ (pc) & $N$ ($\times 10^6$) \\
\mr
A & 2.0 & 200 & 30 \\
B & 2.5 & 100 & 30 \\
C & 2.5 & 200 & 30 \\
D & 2.5 & 300 & 30 \\
E & 3.0 & 200 & 30 \\
F & 2.5 & 500 & 30 \\
G & 2.5 & 500 & 75 \\
\br
\end{tabular}
\end{indented}
\end{table}

\section{The Data Analysis Simulation}
\begin{figure}
\includegraphics[angle=0,width=1\textwidth]{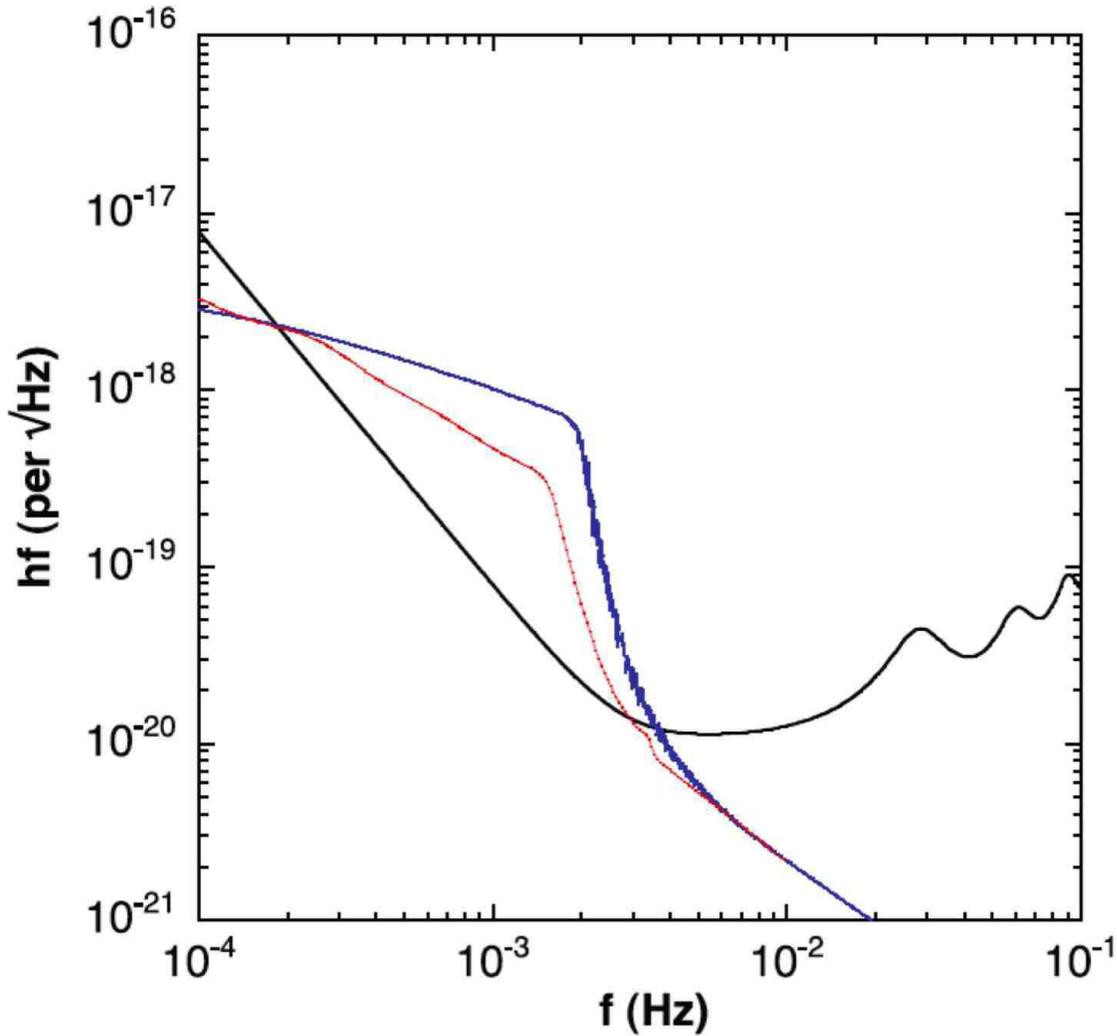}
\caption{\label{background}Sensitivity curve showing the approximated confusion-limited signal for a realisation of model C, along with the standard Hils \etal curve and the LISA sensitivity curve. Our approximated curve is slightly stronger and extends to slightly higher frequencies.}
\end{figure}

We assume that the fundamental criterion for detection of a given binary will be the signal-to-noise ratio (SNR), $\rho$. We determine the SNR for a given binary of frequency $f$ according to:
\begin{equation}
\label{snr}
\rho = \frac{h_0\sqrt{T_{\rm obs}}}{\sqrt{S_{\rm n}(f)}},
\end{equation}
where $T_{\rm obs}$ is the duration of the observation, $S_{\rm n}(f)$ is the power spectral density of the strain noise in the detector, and $h_0$ is the characteristic strain amplitude of the binary at the barycentre. The characteristic strain amplitude of a binary at a distance $D$ is calculated from:
\begin{equation}
h_0 = \frac{G \mathcal{M}}{c^2 D}\left(\frac{G\mathcal{M}\pi f}{c^3}\right)^{2/3},
\end{equation}
where the chirp mass is defined to be:
\begin{equation}
\mathcal{M} = \frac{\left(m_1m_2\right)^{3/5}}{\left(m_1 + m_2\right)^{1/5}}.
\end{equation}
We also assume that $S_{\rm n}(f)$ consists of a combination of instrument noise and the confusion-limited noise of the unresolved binaries. Since we are using an angle-averaged assessment for the calculation of the SNR, we determine the instrumental contribution to simply be the square of LISA sensitivity curve~\cite{larson00}, so that $\sqrt{S_{\rm n}(f)} = h_f(f)$. We approximate the confusion-limited noise contribution to $S_{\rm n}(f)$ by first taking a running mean of the signal and then applying a taper function at higher frequencies where the density of sources becomes less than 1 per resolvable frequency bin. We use the taper function of Hils and Bender~\cite{hils97}. The result of this procedure is shown in Figure~\ref{background} along with the LISA sensitivity curve and the standard Galactic white dwarf binary curve of Hils \etal~\cite{hils90}.
Using the combined instrument noise and approximated confusion-limited noise, we then determine that an individual binary is resolvable if its SNR, as calculated using Equation~\ref{snr}, is greater than 5.

Once we have determined the population of resolved binaries, we then divide this group up into monochromatic and chirping binaries. We ignore possible tidal effects and assume that the frequency evolution of the binary is entirely due to the emission of gravitational radiation, so that:
\begin{equation}
\label{fdot}
\dot{f} = \frac{96}{5}\frac{c^3 f}{G\mathcal{M}}\left(\pi f\frac{G\mathcal{M}}{c^3}\right)^{8/3}.
\end{equation}
If the frequency evolves through at least one resolvable frequency bin during the observation period, we assume that $\dot{f}$ can be recovered from the data analysis and so the mass-distance degeneracy can be broken. Thus, we have obtained two populations of resolved binaries: monochromatic binaries whose sky location can be determined and chirping binaries whose three dimensional position in the Galaxy can be determined. For our models with $N = 30 \times 10^6$ binaries, the number of resolved chirping binaries was $\sim 1300$ and the number of resolved monochromatic binaries was $\sim 25000$. The numbers for the resolved populations of model G scale linearly with the increase in $N$. It is these populations that we will compare to the underlying total Galactic population to determine the selection biases in the observed samples.

\section{Selection Biases}
We look at three different biases in the two populations of resolved systems: the spatial distribution, the frequency distribution, and the chirp mass distribution.

\begin{figure}
\includegraphics[clip=true,width=0.5\textwidth]{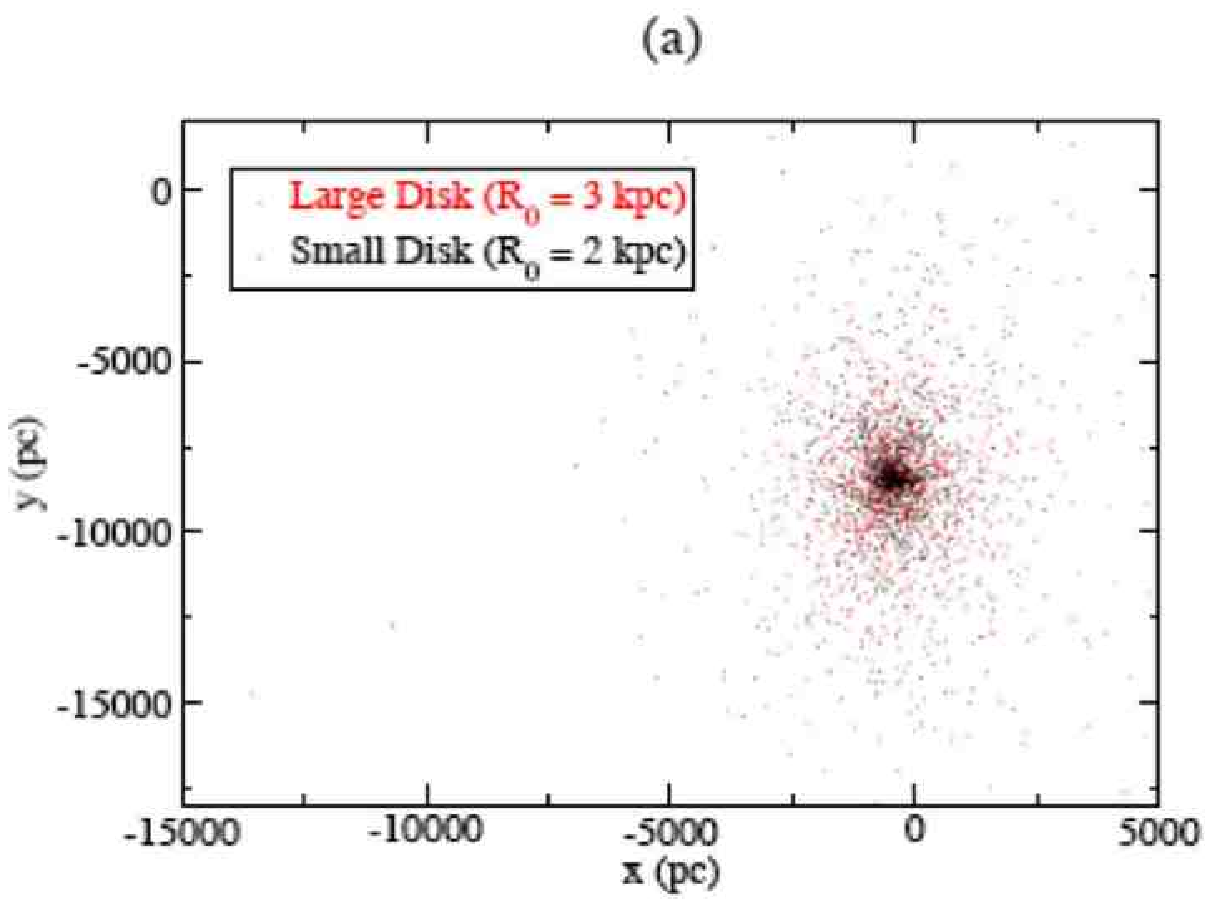}
\includegraphics[clip=true,width=0.5\textwidth]{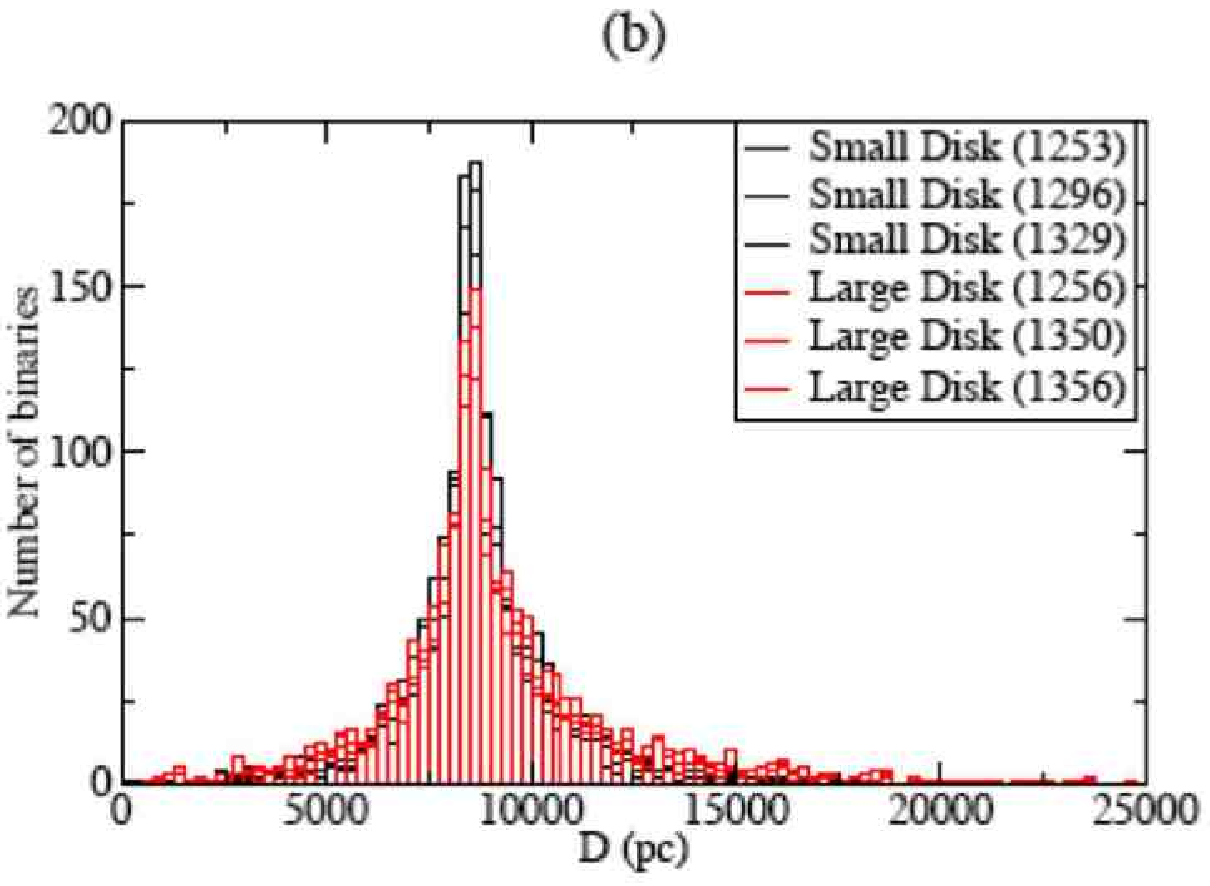}
\caption{\label{chirpspace} Spatial locations of resolved binaries in the Galactic plane for model A (small disk) and model E (large disk) are shown in plot (a). The Sun is located at the origin in this plot. Plot (b) shows a histogram of distance to resolved chirping binaries for 3 realisations of model A and model E.}
\end{figure}

The spatial distribution of chirping binaries is relatively even throughout the Galaxy as can be seen in plot (a) of Figure~\ref{chirpspace} for both the large disk (model E) and small disk (model A) models of the Galaxy. However, as can be seen in plot (b) of Figure~\ref{chirpspace}, it is still possible to distinguish between the different models. This capability is explored in more detail in Larson \etal~\cite{larson07}.

\begin{figure}
\includegraphics[clip=true,width=0.5\textwidth]{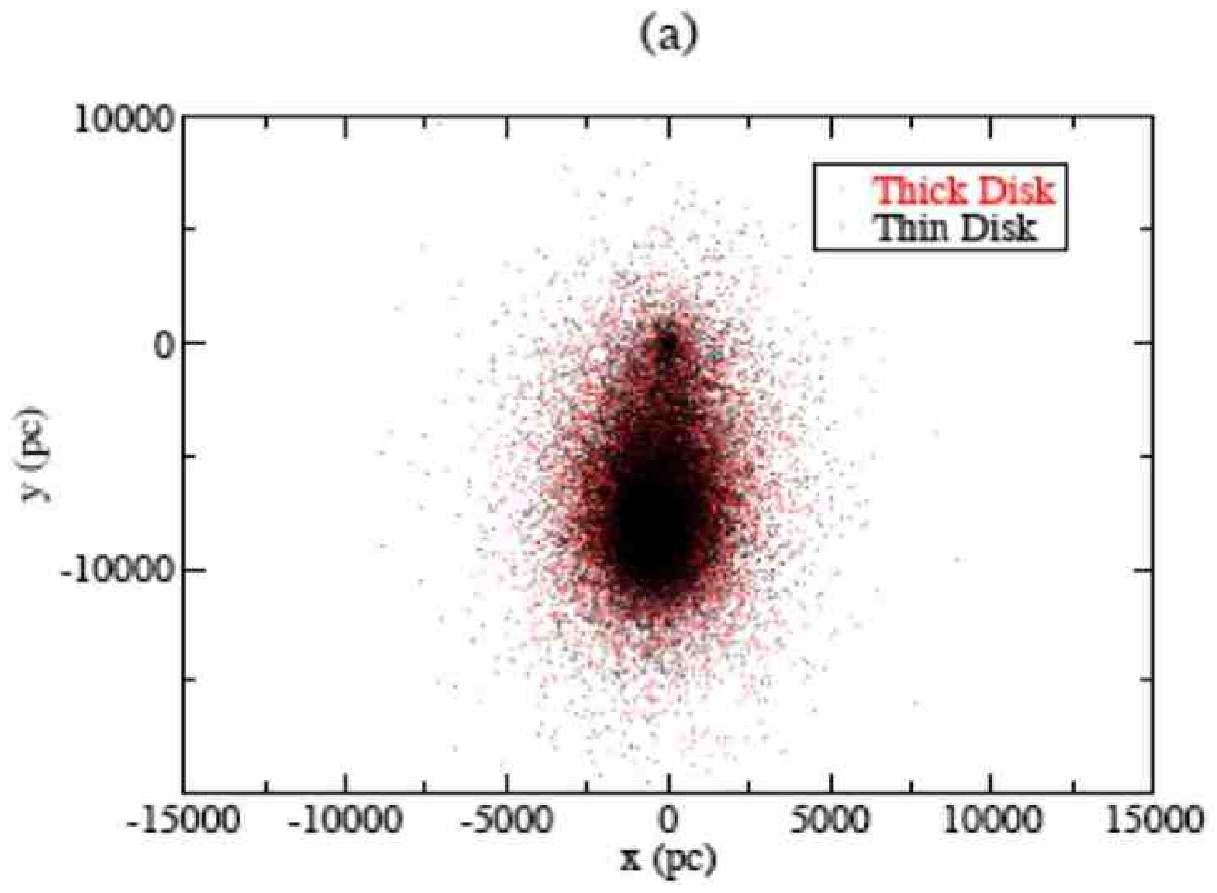}
\includegraphics[clip=true,width=0.5\textwidth]{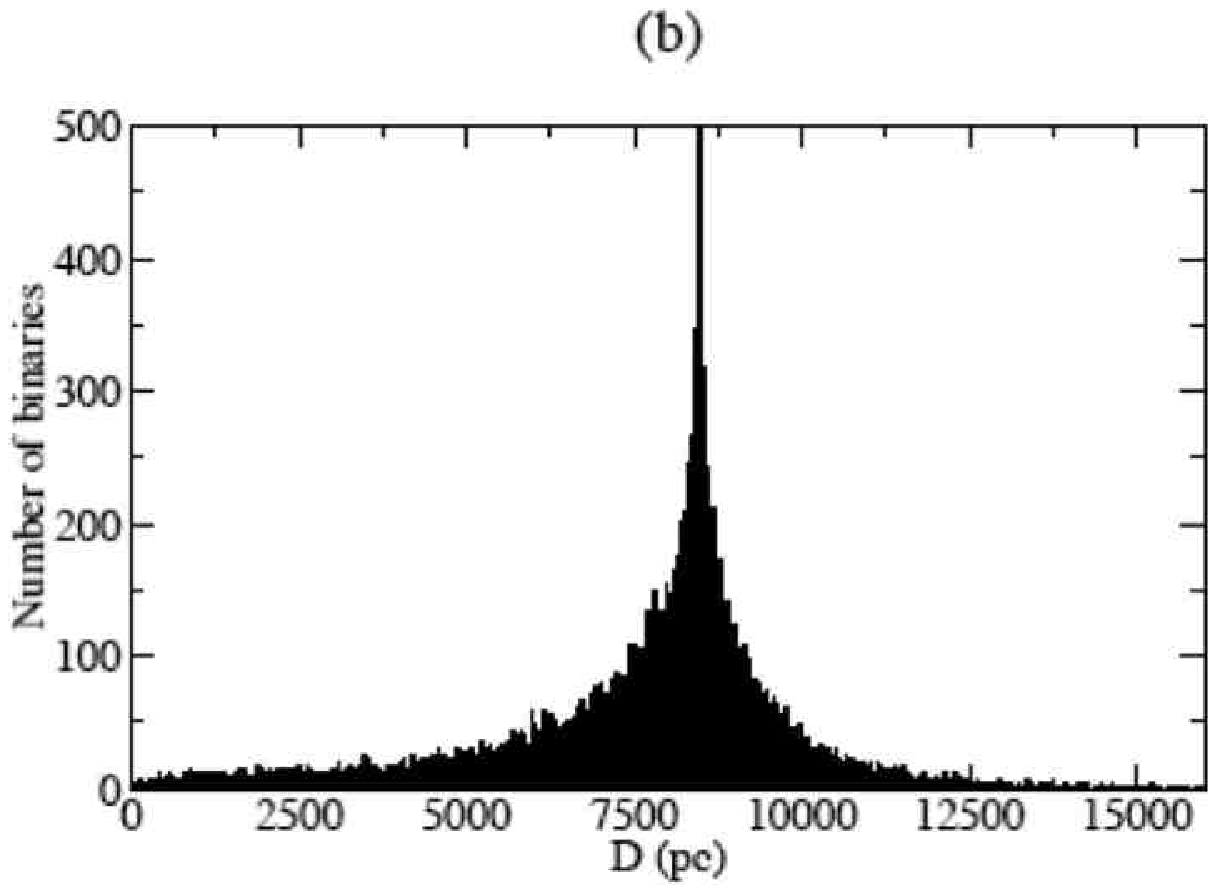}
\caption{\label{monospace} Spatial locations of resolved binaries in the Galactic plane for model B (thin disk) and model F (thick disk) in plot (a). The Sun is located at the origin in this plot. Plot (b) shows a histogram of distance to resolved monochromatic binaries for model C. Note the excess of systems between the Sun and the Galactic center.}
\end{figure}

\begin{figure}
\includegraphics[clip=true,width=1.0\textwidth]{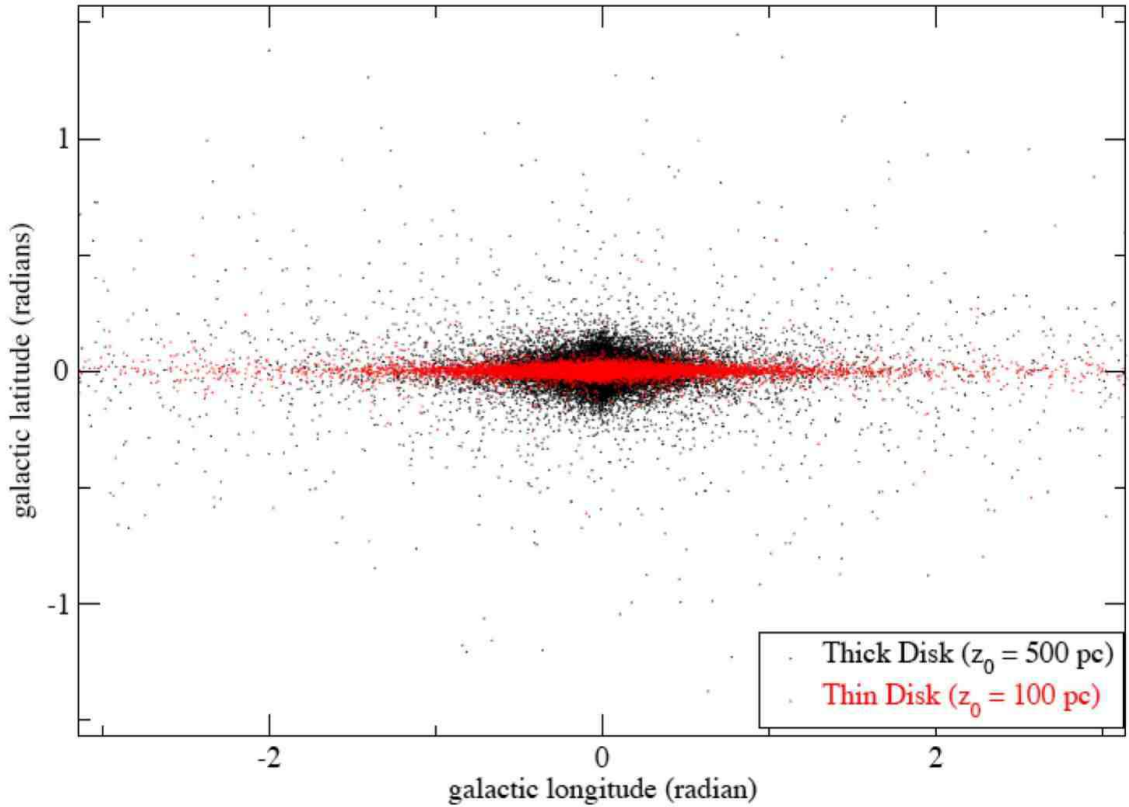}
\caption{\label{scaleheight} Sky locations of resolved monochromatic binaries in Galactic longitude and latitude for model B (think disk) and model F (thick disk).}
\end{figure}

The spatial distribution of monochromatic binaries also samples the entire Galaxy, but there is a slight preference for closer systems as can be seen in the spatial distribution and the histogram of distances shown in Figure~\ref{monospace}. This can be understood by noting that the monochromatic population of binaries includes a higher number of lower frequency binaries. These systems are inherently less luminous than similar binaries at higher frequency. Consequently, they must be closer in order to cross our detection threshold of SNR $\ge 5$.
The monochromatic binaries can also be used to determine the scale height of the Galaxy as shown in Figure~\ref{scaleheight}. Again, this capability will be treated in more detail in Larson \etal~\cite{larson07}.

\begin{figure}
\includegraphics[clip=true,width=0.5\textwidth]{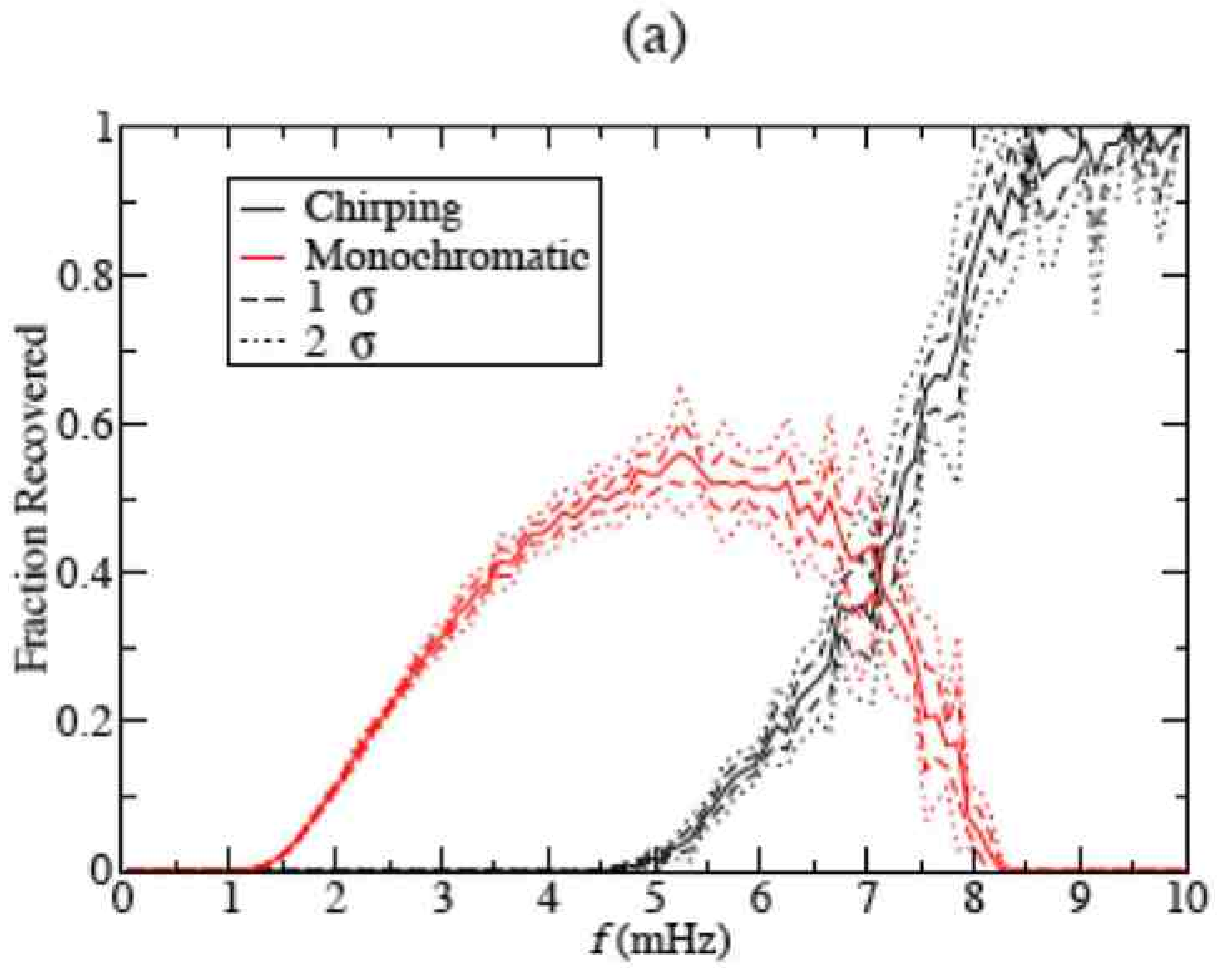}
\includegraphics[clip=true,width=0.5\textwidth]{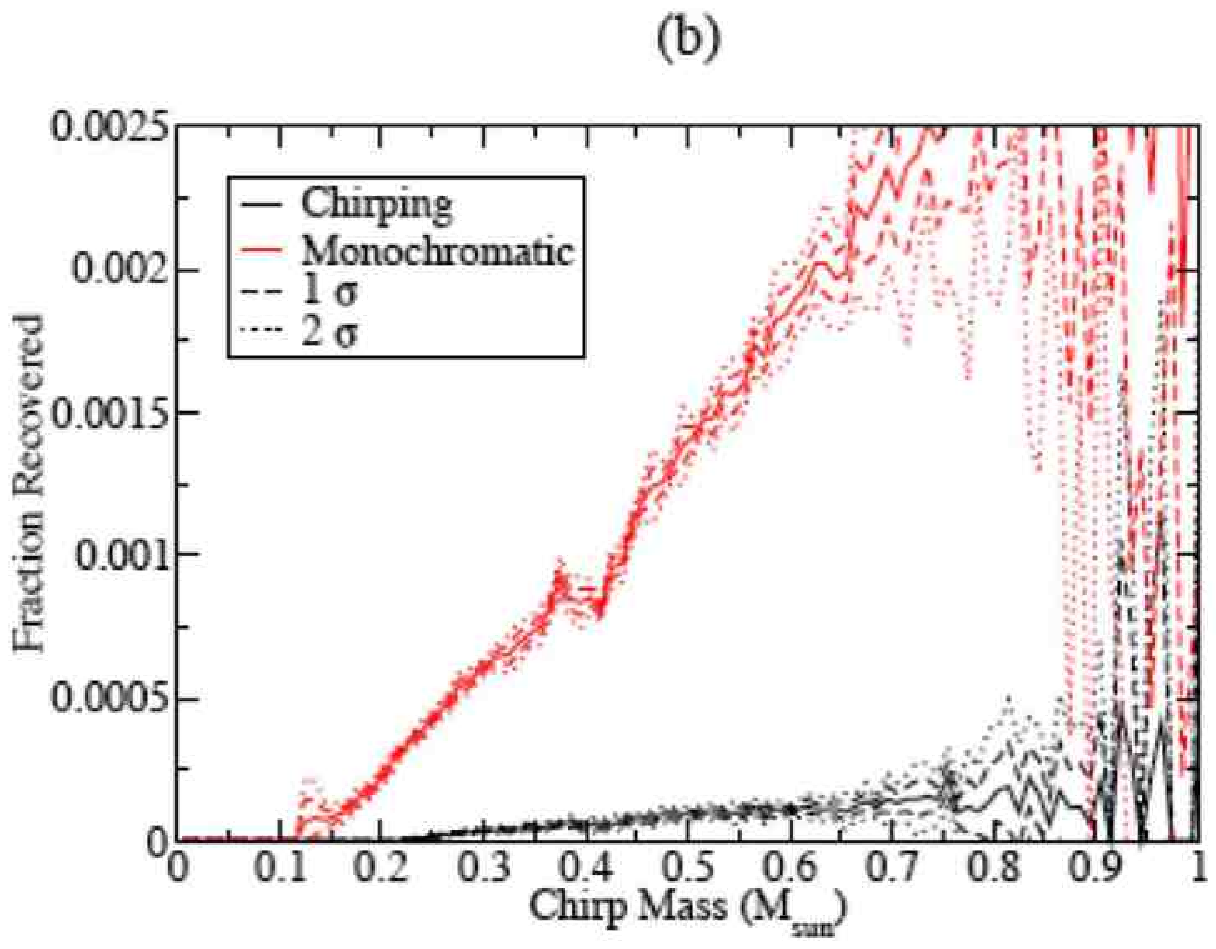}
\caption{\label{biasplots} The recovery percentage for both chirping (black curves) and monochromatic (red curves) binaries. The solid curve represents the mean recovery percentage for all realisations and models. The dashed and dotted curves show the 1 $\sigma$ and 2 $\sigma$ spread in recovery percentages, respectively. Plot (a) shows the bias in frequency and plot (b) shows the bias for chirp masses.}
\end{figure}

The frequency distribution of the resolved chirping binaries is strongly biased towards higher frequencies, with virtually none observed below $\sim 4.5$ mHz. This is partly due to the fact that lower frequency binaries simply will not have a measurable chirp as can be seen by the frequency dependence in Equation~\ref{fdot}. On the other hand, strongly chirping binaries are rare in the Galaxy, and so at frequencies above $\sim 10$ mHz the recovery rate approaches 100\%. This bias is unchanged in all realisations of all of our models. The mean and 1 and 2 $\sigma$ curves of the percentage recovery of chirping binaries for all models are shown as the black curves in plot (a) of Figure~\ref{biasplots}.
For the monochromatic resolved binaries, the bias is again toward higher frequencies, with a negligible fraction of recovered binaries having frequencies below $\sim 1$ mHz. The recovery percentage increases to $\sim 50\%$ at $\sim 5$ mHz and then drops down to zero again above $\sim 8$ mHz. The reduction in recovery rates for monochromatic binaries at higher frequencies is a direct result of the fact that very few binaries at these frequencies will not be chirping. Thus, the set of recovered monochromatic binaries is not sampling the Galactic population of binaries above about 8 mHz. The mean and 1 and 2 $\sigma$ curves of the percentage recovery of monochromatic binaries for all models are shown as the red curves in plot (a) of Figure~\ref{biasplots}.

The distribution in chirp masses for the resolved chirping binaries rises from a negligible recovery rate starting at $\mathcal{M} \sim 0.2~{\rm M}_\odot$. The rise is very shallow throughout and slightly less steep at chirp masses above $\mathcal \sim 0.5~{\rm M}_\odot$. The recovery rate can be seen in plot (b) of Figure~\ref{biasplots}. When we look at the recovery rate for the resolved monochromatic binaries, we see a steeper rise in recovery that starts at a lower chirp mass of $\mathcal{M} \sim 0.1~{\rm M}_\odot$. The recovery rate never flattens out above any chirp mass. These recovery rates are shown in plot (b) in Figure~\ref{biasplots}. The overall increased recovery rate from monochromatic binaries simply stems from the fact that there are more monochromatic binaries recovered than there are chirping binaries. However, it is the difference between the properties of chirping binaries relative to the monochromatic binaries that explains the difference in the shapes of the curves. The very low chirp mass binaries that are recovered in the population of monochromatic binaries are simply too weak to significantly chirp, and so these will not be observable in the chirping population. The steady rise in the fraction of the recovered monochromatic binaries is a result of the fact that an increased chirp mass leads to a larger amplitude gravitational wave. Consequently, binaries of a given frequency but larger chirp mass can be observed at greater distances. This means that a larger spatial sample of the Galactic population can be observed at larger chirp mass. However, there is an additional effect due to the fact that a larger chirp mass allows for the detection of binaries at lower frequencies as well. Since the population of Galactic binaries grows significantly with decreasing frequency, the increase in chirp mass also allows for the detection of a larger fraction of the Galactic population. These two effects combine for the steady rise in the recovered fraction of monochromatic binaries with chirp mass.

\section{Conclusions}
We have simulated a number of realisations of galaxies with a limited range of structural parameters in order to test which components of the Galactic population of detached white dwarf binaries contribute to the population of resolved binaries that will be used to achieve LISA's science goals of surveying compact stellar mass binaries and studying the structure of the Galaxy. In particular, we have looked at the selection biases in spatial distribution throughout the Galaxy, the frequency distribution, and the chirp mass distribution. We have separated the resolved population into monochromatic binaries (which can yield sky location information) and chirping binaries (which can yield distance information and white dwarf structure information). For both monochromatic and chirping resolved populations, the sample is drawn from the entire spatial distribution of the Galaxy. There is a slight excess of nearby systems in the monochromatic population. There is no discernible bias in the spatial distribution of the chirping binaries that may be used to uncover the three-dimensional structure of the Galaxy.

Both populations exhibit biases in frequency and chirp mass. The density of binaries in frequency space increases dramatically with decreasing frequency, so the contribution of confusion-limited noise significantly increases at lower frequencies. Consequently, the detection efficiency is much greater at the higher frequencies where the noise is dominated by instrument noise. Thus, the population of resolved binaries are drawn almost exclusively from systems with frequencies above 1 mHz. For chirping binaries the low frequency cut-off is at 4 mHz. Above 8 mHz, nearly all the Galactic binaries will have a measurable chirp, and consequently the monochromatic population contains almost no binaries above this frequency. The chirping population has nearly a 100\% recovery rate above this frequency. Therefore, when the chirping population is used to map out the structure of the Galaxy, it is actually mapping out the structure of the high frequency population of binaries. Although the frequency cut-off for monochromatic binaries is lower at 1 mHz, this is still only a sampling of the very small percentage of the Galactic binary population. Thus, measurements of the disk scale height that can be made using the sky position of the monochromatic binaries, will be tracing the scale height of the binaries with gravitational wave frequencies above 1 mHz.

The bias towards higher frequencies in both the monochromatic and chirping populations contributes to a bias towards higher chirp mass as well since lower mass binaries will evolve to Roche lobe overflow at lower frequencies. This is evident in the mass distributions of the resolved populations of both monochromatic and chirping binaries. There is an additional bias towards higher chirp mass due to the fact that these systems will also have larger amplitudes and thus will be detectable at greater distances. In fact, one of the more remarkable features of both populations of resolved binaries is that they show only a small bias towards nearby systems. This implies that the resolvable systems are rare but strong systems. Thus, the population of resolved binaries for LISA will be able to shed light on the Galactic structure of the high-mass, high-frequency population of close white dwarf binaries.

\end{document}